\documentclass{elsart}

\usepackage{graphicx}

\usepackage{amssymb}
\usepackage{url}

\newcommand{\sd}{\mathrm{d}}

\newcommand{\ex}[1]{\mathrm{e}^{#1}}

\newcommand{\isotope}[2]{^{#1}\textrm{#2}}

\begin{document}

\begin{frontmatter}


\title{Digital pulse-shape discrimination of
  fast neutrons and $\gamma$ rays}
\author[label1]{P.-A. S{\"o}derstr{\"o}m\corauthref{cor1}},
\ead{Par-Anders.Soderstrom@fysast.uu.se}
\author{J. Nyberg\thanksref{label1}},
\ead{Johan.Nyberg@fysast.uu.se}
\author{R. Wolters\thanksref{label1}}
\address[label1]{Department of Physics and Astronomy, Uppsala
  University, SE-75121 Uppsala, Sweden}
\corauth[cor1]{Corresponding author}





\begin{abstract}
Discrimination of the detection of fast neutrons and $\gamma$ rays in
a liquid scintillator detector has been
investigated using digital pulse-processing techniques.
An experimental setup with a $^{252}$Cf source, a BC-501 liquid
scintillator detector, and a BaF$_2$ detector was used to
collect waveforms with a 100 Ms/s, 14 bit sampling ADC.
Three identical ADC's were combined to increase the sampling frequency
to 300 Ms/s. Four different digital
  pulse-shape analysis algorithms were developed and compared to each
  other and to data obtained with an analogue neutron-$\gamma$
  discrimination unit. Two of the digital algorithms were based on the
  charge comparison method, while the analogue unit and the other two
  digital algorithms were based on the zero-crossover method. Two
  different figure-of-merit parameters, which quantify the
  neutron-$\gamma$ discrimination properties, were evaluated for all
  four digital algorithms and for the analogue data set. All of the
  digital algorithms gave similar or better figure-of-merit values
  than what was obtained with the analogue setup.
  A detailed study of the discrimination properties as a function of
  sampling frequency and bit resolution of the ADC was performed. It was shown that a sampling ADC with a bit
  resolution of 12 bits and a sampling frequency of 100 Ms/s is
  adequate for achieving an optimal neutron-$\gamma$ discrimination
  for pulses having a dynamic range for deposited neutron energies of
  $0.3$-$12$ MeV. An investigation of the influence of the sampling frequency on the
time resolution was made. A FWHM of 1.7 ns was obtained at 100 Ms/s.
\end{abstract}

\begin{keyword}
  Digital pulse-shape discrimination \sep fast-neutron detection \sep
  liquid scintillator \sep BC-501 \sep sampling ADC
\PACS 
\end{keyword}
\end{frontmatter}

\section{Introduction}

In experimental studies of the structure of exotic nuclides far from the line of $\beta$ stability, it is important to accurately and efficiently identify the residual nuclides produced in the
nuclear reactions. One of the most common type of reactions used is heavy-ion fusion-evaporation, in which two nuclei fuse to form a compound nucleus, which decays by evaporating a number of light particles, mainly neutrons, protons, and/or $\alpha$
particles.

The exotic nuclides of interest are produced with very small cross
sections in this type of reactions. A usual method of identifying
the proton ($Z$) and neutron ($N$) number of these nuclides, is to detect 
the type and number of emitted light particles.
If all emitted particles are detected in each reaction,
the final nuclide can uniquely be identified by subtracting 
the sum of the $Z$ and $N$ of the emitted 
particles from the $Z$ and $N$ of the compound nucleus, 
With careful analysis of the distributions of the number of detected
light particles, it is possible to identify the $Z$ and $N$ of the
residual nuclides even if all emitted particles are not detected in
every reaction, which usually is the case.

The light charged particles are usually detected by a highly efficient
Si or CsI detector array, while the neutrons are detected by an array
of liquid scintillator detectors.
The requested nuclear structure information is obtained
by using a high-resolution $\gamma$-ray spectrometer for detection
of the emitted $\gamma$ rays in coincidence with the light particles.

A clean and efficient detection of the number of emitted neutrons in
each reaction, is of utmost importance for studies of nuclides located
close to the proton-drip line. The compound nuclei created in such
studies decay mainly by emission of a number of protons, which brings
the $Z$ of the final nucleus closer to the $\beta$-stability line. Only
very rarely are neutrons emitted and by detecting them one can
identify the neutron number of the produced rare exotic nuclides. One
of the major challenges of the detection of these rare neutrons, is to
discriminate between neutrons and other particles, mainly $\gamma$
rays, which are also registered by the neutron detectors.

Future studies of exotic nuclei will mainly be performed by using  
reactions induced by radioactive instead of stable heavy ions. This allows for
a production of compound nuclei located even further from the line of
$\beta$ stability. Experiments with radioactive ion beams are often
hampered by the high $\gamma$-ray background radiation originating from
the decay of the radioactive beam. Neutron detector
arrays, to be used in future experiments with high-intensity
radioactive ion beams, for example at the HISPEC \cite{2006IJMPE..15.1967P} and DESPEC \cite{2006IJMPE..15.1979R} setups at NuSTAR/FAIR \cite{npn_nustar, 2007JPhG...34..551H} and at SPIRAL-2 \cite{2007AIPC..891...91L}, must therefore be designed to cope with such a
highly increased $\gamma$-ray background. Another future application
is experiments with extremely high-intensity stable beams,
which also generate intense background radiation, leading to
problems of clean neutron detection due to random and pileup effects.

The need for efficient neutron detection is not only required in
heavy-ion fusion-evaporation reactions close to the proton drip-line,
but also e.g. as ``veto'' detectors in studies of neutron-rich nuclei and
in neutron spectroscopy and neutron correlation studies of
neutron-halo nuclei.

This paper presents a study of the discrimination of the detection of
neutrons, with energies from about 0.3 MeV to 10 MeV, from $\gamma$
rays in a liquid scintillator detector using a fast sampling
analogue-to-digital converter (ADC) and digital pulse-processing techniques. Several different digital neutron-$\gamma$ discrimination (NGD)
algorithms were implemented and the discrimination quality was
compared with what could be achieved with an analogue system. For the
digital versions, the discrimination properties were studied as a
function of sampling frequency and bit resolution of the ADC. The main
focus of the study is on the detection of pulses corresponding to low
energy deposition in the detector, since in this case the
NGD is more difficult and the needs for
improvements are larger.

Section \ref{sec:liq} is a brief introduction to the principle of
pulse-shape discrimination in liquid scintillators.
The details of the performed experiment are presented in section 
\ref{sec:experiment}, the implemented digital pulse-shape
algorithms in section \ref{sec:disc}, and the analysis and results in
section \ref{sec:analysis}.

Preliminary results of this work have been published in \cite{varenna}.

\section{Liquid scintillators and pulse-shape discrimination} \label{sec:liq}

In order to get a good and clean detection of fast neutrons, several
features of the detector design needs to be taken into
consideration. The organic liquid scintillator, which is a so called
proton recoil detector, has been used very successfully in previous
neutron detector arrays \cite{1991NIMPA.300..303A,1999NIMPA.421..531S,2004NIMPA.530..473S}. This is partly because of its good
efficiency for detection of fast neutrons, which is due to the large
cross section for elastic neutron-proton scattering, and partly
because of its excellent pulse-shape discrimination (PSD) properties,
which allows for a discrimination of neutrons and $\gamma$ rays.

A possible problem of using a detector, which is based on elastic
neutron-proton scattering, is that the neutron will have a substantial
kinetic energy after the interaction and a velocity in some random
direction. In a closely-packed detector array, multiple scattering
between different segments becomes a problem, which can cause quite
severe errors in counting the number of neutrons emitted in each
reaction. Methods to correct for this have been developed with good
results \cite{1997NIMPA.385..166C,2004NIMPA.528..741}. This requires,
however, very good discrimination of neutrons and $\gamma$ rays, since
it has been shown that even a small amount of $\gamma$ rays
mis-interpreted as neutrons dramatically reduces the quality of the
multiple-scattering rejection \cite{2004NIMPA.528..741}.

\begin{figure}[ht] 
  \centering
\begin{center}
\includegraphics[width=\columnwidth,bb=0 0 567 283]{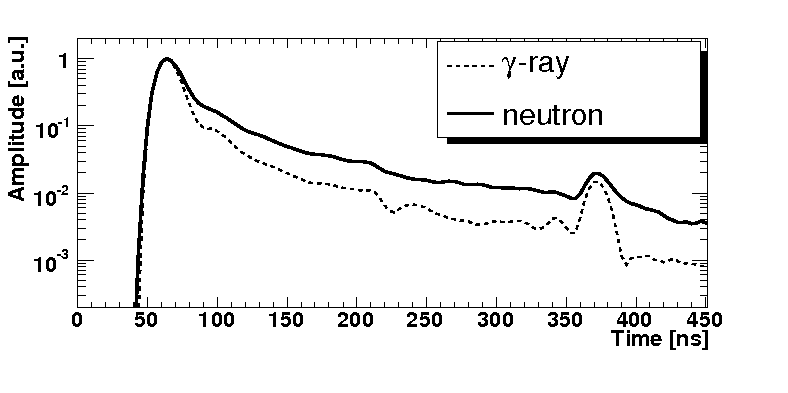}
\end{center}
  \caption{Pulse shapes from a BC-501 liquid scintillator detector measured with a fast sampling ADC, as described in section~\ref{sec:experiment}. The pulses are averaged over about 70~000 $\gamma$-ray pulses and 40~000 neutron pulses and normalized to 1.\label{fig:ngamma}}
\end{figure}

The problem of discriminating between neutron and $\gamma$-ray
interactions in liquid scintillators by using analogue electronics is
well studied. Several methods exist to accomplish this, for example by
measuring the zero-crossover (ZCO) time of a shaped pulse
\cite{alexander1961,roush1964psd} or by comparing the charge collected
from different parts of the pulse \cite{brooks}. The performance of
these methods with respect to each other is also well known in the
analogue case \cite{1995NIMPA.360..584W}. With fast sampling ADC's
becoming better and available at smaller costs, new advanced
algorithms for NGD have been developed and
used with very good results. A correlations approach \cite{2003NIMPA.497..467K}, a curve fitting method \cite{2002NIMPA.490..299T} and a pulse gradient method \cite{2007NIMPA.578..191D,2007NIMPA.583..432A} have for example all yielded good results regarding the discrimination of neutrons and $\gamma$ rays.

\section{Experiment\label{sec:experiment}}

The experiment was carried out using a liquid scintillator detector
from the NORDBALL neutron detector array \cite{1991NIMPA.300..303A}, a
BaF$_2$ detector for triggering and time reference, and a
$\isotope{252}{Cf}$ source.

The scintillator liquid of the neutron detector was of type BC-501 \cite{bc501a}.
The liquid was contained in a 3.3 litre hexagonal aluminium can and it
was viewed through a glass window by a 14-stage, 5'' diameter
PMT of type Philips XP2041. This PMT is
designed to run optimally at about 2200 V. In this experiment it was
run at 1750 V, in order to increase the energy range of the pulses
from the anode output which could be sampled by the ADC without being
saturated. The trigger and time reference detector was a 2''${\times}$2'' BaF$_2$
scintillator with a Philips XP2020Q PMT.

The $\isotope{252}{Cf}$ source consisted of several weak sources
placed in a plastic container of cylindrical shape (diameter 1.5 cm,
length 6 cm, wall thickness 2 mm). The radioactive material of the
source was distributed over a volume of about 1 cm$^3$. The source
had a total activity of about 200 kBq, which implies an emission of
about $2 \cdot 10^4$ neutrons and $6 \cdot 10^4$ $\gamma$ rays per second. 
The source was placed between the two detectors, with a distance of 2 cm
and 75 cm from the front of the BaF$_2$ and BC-501 detectors,
respectively.

\begin{figure*}
\centering
\includegraphics[width=\textwidth,bb=0 0 842 595]{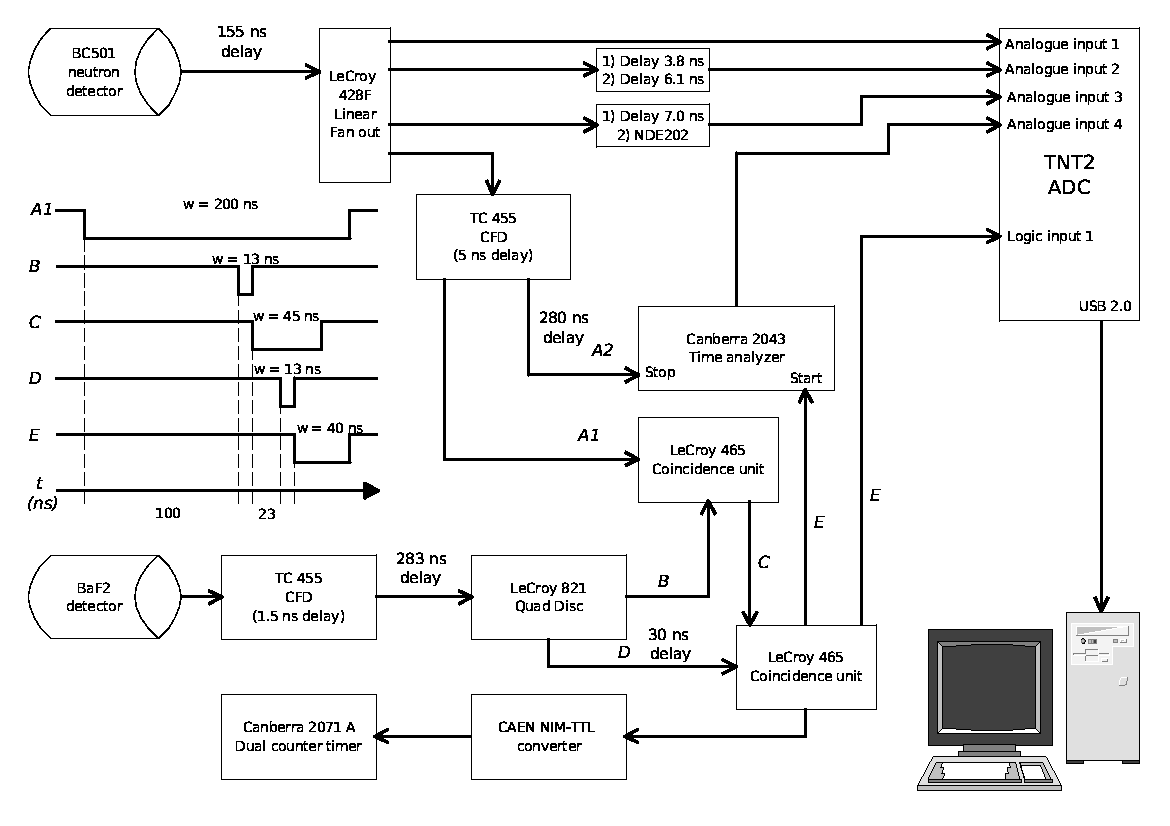}
\caption{    Schematic electronics block and timing diagrams of the setup
    used in this work. The delay units consist of coaxial cables of type
    RG58 (delays $>10$ ns) or RG174. The signals shown in the timing
    diagram correspond to the detection of prompt $\gamma$ rays in the
    two detectors.
    1) 300 Ms/s and
    2) 200 Ms/s setup.
\label{fig:setup}}
\label{fig:electronics}
\end{figure*}

Schematic block and timing diagrams of the electronics setup are shown
in fig.~\ref{fig:setup}. The anode output of the neutron detector was
connected to a LeCroy 428F linear fan-out (LFO). A reflection of the
signal between the anode output and the LFO was observed. This effect
was due to a non-optimal design of the voltage divider regarding
impedance matching and it led to a distortion of the signal pulse
shapes. This problem was solved by introducing a $\approx 30$ m long
RG58 coaxial cable between the anode output and the LFO input. The
reflection and distorted part of the signal was thus delayed by about
300 ns and appeared outside the range used for the
pulse-shape analysis. The reflection is clearly seen in
fig. \ref{fig:ngamma} in the time range 360 to 390 ns.  The long RG58
cable also made the conditions more similar to real physics
experiment, in which the detector and the electronics usually are
separated by large distances. This caused an increase in the pulse rise time (10~\% to 90~\%), from $~6.5$~ns before the cable, to $~7$~ns afterwards.

One of the LFO outputs and the anode signal of the BaF$_2$ detector
were sent to constant fraction discriminators (CFD) of type TC
455. The threshold of the BC-501 CFD was set to 22 keV for electrons which
corresponds to 310 keV for recoil protons. The
threshold of the BaF$_2$ CFD was set to a rather high value to reject
signals from X~rays, low energy $\gamma$ rays and noise.

The outputs of the two CFD units were sent to the inputs of a LeCroy
465 coincidence unit (signals A and B in fig. \ref{fig:setup}), in
which an overlap coincidence between the signals in the BC-501 and
BaF$_2$ detectors was created. The second LeCroy 465 unit in
fig. \ref{fig:setup} (signal inputs C and D) made sure that the
leading edge of the output signal E always was determined by the
BaF$_2$ detector. Signal E was finally used as a trigger of the
sampling ADC (logic input 1) and as a start of the time-to-amplitude
converter (TAC). The TAC was stopped by the delayed BC-501 signal A2
and measured the time-of-flight (TOF) difference between the detected
$\gamma$ rays and neutrons in the detectors.

The anode signal of the BC-501 detector was digitized by a TNT2 ADC
unit \cite{TNT2}. The TNT2 is a single width NIM unit and contains
four independent channels, each of which has a 14-bit flash ADC with a
sampling frequency of 100 megasamples per second (Ms/s), an analogue
bandwidth of 40 MHz, and an input range of $\pm 0.62$ V when
terminated in $50 \Omega$. The TNT2 is set up, controlled and read out
by a Linux PC via a USB 2.0 interface. A Java graphical user interface 
is used for control and monitoring of the unit. In this experiment the 
TNT2 was used in the digital oscilloscope mode, which allows for
readout of the digitized waveforms of each of the four channels.

In order to increase the sampling frequency of the ADC from 100 Ms/s
to 300 Ms/s, three of the output signals of the LFO were used for
pulse sampling. One of the outputs was defined as the
zero-delay branch and was connected directly from the LFO to analogue
input 1 of the TNT2. 
The other two outputs were delayed 
relative to the zero-delay branch using short
RG174 cables, before they were fed into analogue inputs 2 and 3 of the
TNT2. The total effective delays (cables and internal delays of the
LFO and TNT2) of analogue inputs 2 and 3 relative to input 1 were 
3.8 ns and 7.0 ns, respectively. 
This was sufficiently close to the 3.33 ns time difference
between samples that would be the result a true 300~Ms/s ADC. All three
sampling channels were gain and time matched against each other
using a 1 MHz sine output from an Agilent function generator of model
33250A. 

\begin{figure} [ht]
  \centering
\includegraphics[width=\columnwidth,bb=0 0 567 283]{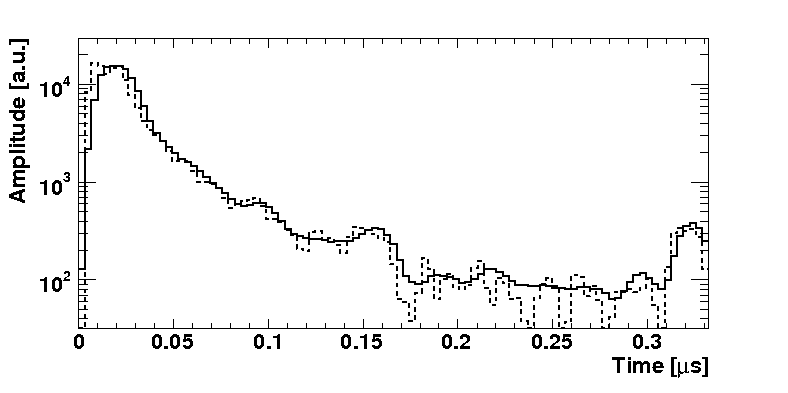}
  \caption{Effects of the limited analogue bandwidth. A typical
    measured pulse shape (solid histogram) and the corresponding
    undistorted pulse shape after correcting for the finite analogue
    bandwidth (dashed histogram) are shown. To remove
    artefacts due to the finite size of the discrete Fourier
    transform, the histograms are smoothed twice using the 353QH
    algorithm \cite{353QH}.\label{fig:analogue}}
\end{figure}

Since the above procedure only effected the digital bandwidth of the
sampling system, one of its consequences was that the analogue
bandwidth did not match the digital bandwidth. To examine the effects
of this, the frequency response of the system was measured. It was
found to have a low pass characteristics of order 2.27 and a cut-off
frequency of 35.7 MHz. The effect of the analogue bandwidth on the
pulse was examined by estimating the shape of an undistorted pulse by
correcting a discrete Fourier transformation with the measured
frequency response. A comparison between a typical pulse and an
estimate of the undistorted version of the same pulse is shown in
fig. \ref{fig:analogue}. An increase in the rise time (10~\% to 90~\%) to $~10$~ns for the digitized pulses was
observed. In the part of the pulse that is most relevant to this work,
the region of the fast and slow decay components, the two pulses
follow each other very closely. The effects on the rise time, and consequently on the fast component, gives an added amplitude uncertainty of $\lesssim 5$~\%.
The effects due to the mis-match of the analogue and digital
bandwidths did not influence the results obtained in this work.

By utilizing well known analogue PSD electronics, a reference
data set was obtained, which could be used for comparisons with the
results obtained by the digital PSD algorithms. The analogue PSD unit
used in the present work was of type BARTEK NDE202 \cite{NDE202},
which was designed and built for the EUROBALL Neutron Wall
\cite{1999NIMPA.421..531S}. This unit has a built in circuit for
NGD based on the ZCO technique. The ZCO
information is available from the NDE202 as a TAC output signal, which
gives the time difference between the leading edge of the signal,
obtained from its internal CFD, and the ZCO signal. The
threshold of the internal CFD was set to 68~keV for electrons, corresponding to 600~keV for protons, i.e. to a higher
value than the TC 455 CFD used by the BC-501 detector. The ZCO TAC
output signal of the NDE202 was sent to analogue input 3 of the TNT2
ADC after being attenuated to be within the amplitude range of the ADC
(see fig. \ref{fig:setup}).

In the experiments which included the analogue PSD unit, only two of the
TNT2 analogue inputs could be used for digitization of the BC-501 anode
signals, as seen in fig.~\ref{fig:setup}. Identical to what was described above, the LFO output for the
zero-delay branch was sent directly to analogue input 1 of the
TNT2. Another LFO output was delayed by 6.1 ns relative to the
zero-delay branch. In this way a sampling frequency of 200 Ms/s could be achieved.

During the experiments the singles rates were about 2 and 30 kHz for
the BC-501 and BaF$_2$ detectors, respectively. The true and random coincidence rates
were about 300~Hz and 0.5~Hz, respectively.

In the experiment two data sets were collected,
one with the 300~Ms/s and the other with the 
200~Ms/s setup, each containing about 6 million events.
The events contained the digitized waveforms of all four channels of
the TNT2. The read out waveforms had a length of 
20 $\mu$s (2000 sampling points) and a starting time about 1.8 $\mu$s 
before the leading edge of the pulses.

A rough energy calibration was
obtained by using a $\isotope{137}{Cs}$ source. The Compton edge at
480~keV was observed by an oscilloscope at the output of the LFO to have an amplitude of about
0.5 V. Using a sine signal with an amplitude of 500 mV from the
Agilent function generator, the ADC could be 
calibrated with a $\gamma$-ray calibration coefficient
$C_{\mathrm{e}}$ that converts from pulse amplitude $A$, measured in channel
numbers by the ADC, to electron energy $E_{\mathrm{e}}$ in keV:
\begin{equation} \label{eq:ce}
E_{\mathrm{e}} = C_{\mathrm{e}} \cdot A, \;\; C_{\mathrm{e}} = 0.076 \; \textrm{channels/keV.} 
\end{equation}

\begin{figure}  [ht] 
  \centering
\includegraphics[width=\columnwidth,bb=0 0 567 405]{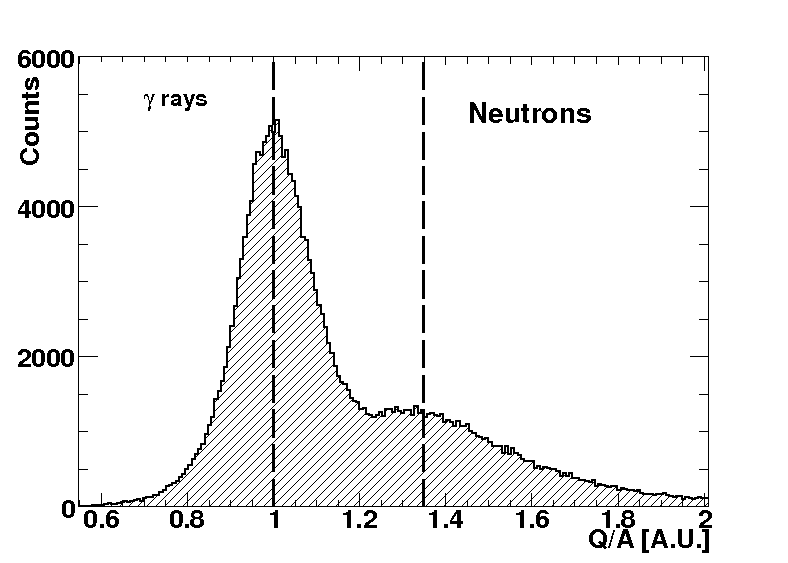}
  \caption{The ratio of total charge to amplitude for a large number
    of waveforms of varying amplitude. The ratio for the peak
    corresponding to $\gamma$ rays is normalized to unity, which gives
    a value of 1.35 for the neutron peak.\label{fig:amptocharge}}
\end{figure}

This amplitude-energy conversion is only valid for $\gamma$-ray
interactions. The following procedure was followed to get an
amplitude-energy conversion for neutron interactions. 
The number of created scintillation photons per deposited energy
is much smaller for proton recoils compared to electrons. The conversion 
from deposited proton recoil energy $E_{\mathrm{p}}$ in MeV to equivalent  
electron energy $E_{\mathrm{ee}}$ in MeV was done by using the relation \cite{epenergy}
\begin{equation} \label{eq:kevee}
  E_{\mathrm{ee}} = 0.83 \cdot E_{\mathrm{p}} - 2.82 \cdot 
  \left(1 - \ex{-0.25 \cdot E_{\mathrm{p}}^{0.93}}\right)\textrm{.}
\end{equation}
This relation holds for the total integrated charge of the measured pulse,
which is proportional to the sum of all created scintillation
photons. The total integrated charge of the pulse is proportional to
the ``area'' between the digitized waveform and its baseline, which can be 
obtained by summing the content of each channel over the whole pulse.
An additional correction was made for conversion from
pulse amplitude to energy for the neutron interactions.
A distribution of the ratio of the total integrated area ($\sim$
total charge) to the amplitude for a large number of pulses with varying
amplitudes is shown in fig. \ref{fig:amptocharge}. 
Two peaks, corresponding to $\gamma$ rays and neutrons are clearly
visible in this plot. The ratio of the position of these peaks is
$k=1.35$, which gives the factor by which a measured amplitude $A$ in
channel numbers should be converted to electron equivalent energy 
$E_{\mathrm{ee}}$ for a neutron interaction:
\begin{equation} \label{eq:etop}
  E_{\mathrm{ee}} = C_{\mathrm{p}} \cdot A = k \cdot C_{\mathrm{e}} \cdot A = k \cdot E_{\mathrm{e}}, \;\; k =\frac{C_{\mathrm{p}}}{C_{\mathrm{e}}}
  \textrm{.}
\end{equation}
A conversion from measured amplitude $A$ to proton recoil energy $E_{\mathrm{p}}$
can finally be done by combining equations \ref{eq:kevee} and \ref{eq:etop} 
and solving numerically for $E_{\mathrm{p}}$. All amplitudes given in keV in the
rest of this paper correspond to $E_{\mathrm{e}}$ and thus have to be
translated to $E_{\mathrm{p}}$ according to this prescription.

A gain and time matching of the analogue inputs 1-3 of the ADC was
done to correct for any differences in the inputs. This was achieved by 
feeding a sine wave from the Agilent function generator into the 
LFO and from there into the TNT2. A few thousand calibration events were
collected and a fit of a sine function was made for each channel in
each event. From the fits the relative gain and time between the
three channels were extracted.
The short and long term stability of the setup regarding changes in
the relative gain of the analogue channels of the TNT2 was excellent and
did not need any further corrections.

For the real data, obtained with the $^{252}$Cf source, 
the signal baseline of each read out waveform was calculated 
on an event-by-event basis. This was necessary 
because the LFO introduced a low frequency noise on the
signal waveforms, which was observed as a small instability of the
baseline. The position of the baseline was obtained by taking an
average of 75 sampling points in a region starting 1.8~$\mu$s and ending 
1.1~$\mu$s before the leading edge of the pulse.

\begin{figure}  [ht] 
  \centering
\includegraphics[width=\columnwidth,bb=0 0 567 340]{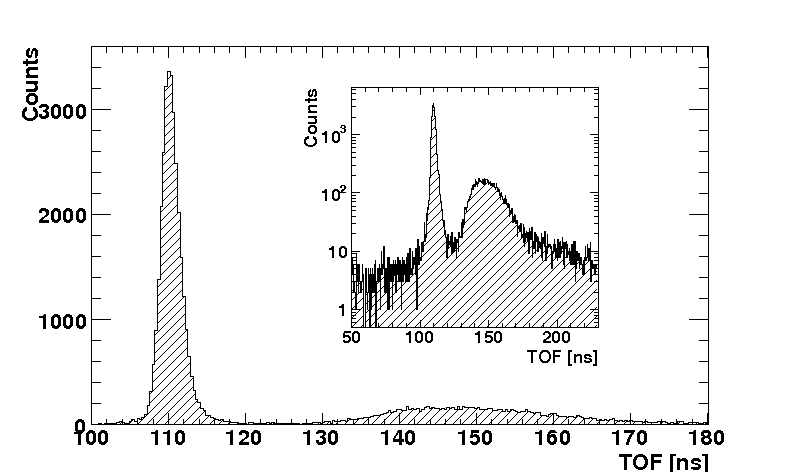}
  \caption{Measured time-of-flight distribution of pulses with amplitude in the range $E_\mathrm{e}=90$-$700$~keV. The $\gamma$-ray peak, to the left, has a FWHM of about $3.1$~ns. The inset shows the TOF distribution in logarithmic scale.\label{fig:TOF}}
\end{figure}

The time information was extracted from the TAC signal. The baseline of this signal was determined for each event as an average before and after the flat top. The flat top itself was then fitted with a constant value, and the time information extracted from this value. See fig.~\ref{fig:TOF} for the obtained TOF distribution.

\section{Digital PSD algorithms for neutron-$\gamma$ discrimination}
\label{sec:disc}

In order to get the requested information from a PSD analysis an
observable, which differs between the interacting particles, is needed.
A general way to define such an observable, $S$, is by
the integral
\begin{equation} \label{eq:psdobs}
 S = \int_{0}^{T} p(t) w(t) \sd t,
\end{equation}
where $T$ is the time of examination of the pulse $p(t)$ and $w(t)$
is a weighting function with the purpose to enhance the features used
for the PSD. The choice of $w(t)$ determines the quality of the discrimination. It has been shown
\cite{gattimartini} that the theoretically optimal form of 
$w(t)$ is
\begin{equation} \label{eq:optimalw}
  w(t) =
  \frac{\bar{n}(t)-\bar{\gamma}(t)}{\bar{n}(t)+\bar{\gamma}(t)},
\end{equation}
where $\bar{n}(t)$ and $\bar{\gamma}(t)$ are the average pulse shapes
for neutron and $\gamma$-ray interactions respectively. The existing
analogue PSD algorithms attempt to create a weighting function $w(t)$
in hardware, which is as close as possible to
eq.~(\ref{eq:optimalw}). This can be done either directly by finding
an analytical expression, which approximates $w(t)$ for the algorithm
in use, or indirectly by comparing this expression with the optimal
weighting function. See for example Ref.~\cite{roush1964psd} for this
kind of analysis of the ZCO method.

The two classes of PSD algorithms implemented digitally in this work
are the well known ZCO and charge comparison (CC) methods. These were
chosen for their simplicity and since they are known to work well both
for analogue and digital PSD systems. The simplicity of the
algorithms also makes the required computing power minimal, which is
an important parameter for future real-time implementations. The implemented algorithms and the analogue discrimination are listed in 
table~\ref{tab:algorithms}.

\begin{table}
  \caption{Classification of the PSD algorithms studied in this paper.}
  \label{tab:algorithms}
  \begin{tabular*}{\columnwidth}{@{\extracolsep{\fill}}lccc}
    \hline
      \textbf{Methods}               &   & Algorithms &  \\
                     & Analogue  & Digital & Digital\\
    \hline
      \textbf{ZCO}     & NDE202 & Convolution & Integrated rise time \\
      \textbf{CC}  & - & Slow component & GDM Integral\\
    \hline
  \end{tabular*}
\end{table}

\subsection{Digital leading-edge discriminator} \label{ss:led}

The starting time of the pulse was determined by a simple digital leading-edge discriminator (LED), which was implemented in the following way. The first sampling point of the recorded waveform with an amplitude value larger than a chosen threshold value was identified. The time at which the pulse crossed the threshold was determined by making a linear interpolation between this sampling point and the previous one. No corrections for time walk were made in this work.

\subsection{ZCO: Analogue and convolution algorithm} \label{ss:afa}

The analogue ZCO method is usually implemented by a
shaping of the detector pulse into a bipolar signal, followed by a
zero-crossing circuit, which detects the time when the amplitude of
the bipolar signal changes polarity, the ZCO time. The
shaping can be done by using double delay lines \cite{alexander1961}
or an RC-CR integrating and differentiating network
\cite{roush1964psd}. The analogue PSD algorithm of the NDE202 unit
used in this work contains a bipolar RC-CR shaping amplifier and a
zero-crossing detector.

In this work, a digital version of the ZCO method, here named the
convolution algorithm, was implemented by imitating a RC-CR network. A
digital filter $h(t)$ of length $t_L$ was applied to the raw input
pulse $p(t)$ to create the shaped pulse $f(t)$ as the convolution
(denoted by $\ast$) of $p(t)$ and $h(t)$:

\begin{equation} \label{eq:ft}
 f(t) = p(t) \ast h(t) \equiv \sum_{\tau=t-t_L}^{t} p(\tau) h(t-\tau).
\end{equation} 

\begin{figure} [ht] 
  \centering
\includegraphics[width=\columnwidth,bb=0 0 567 567]{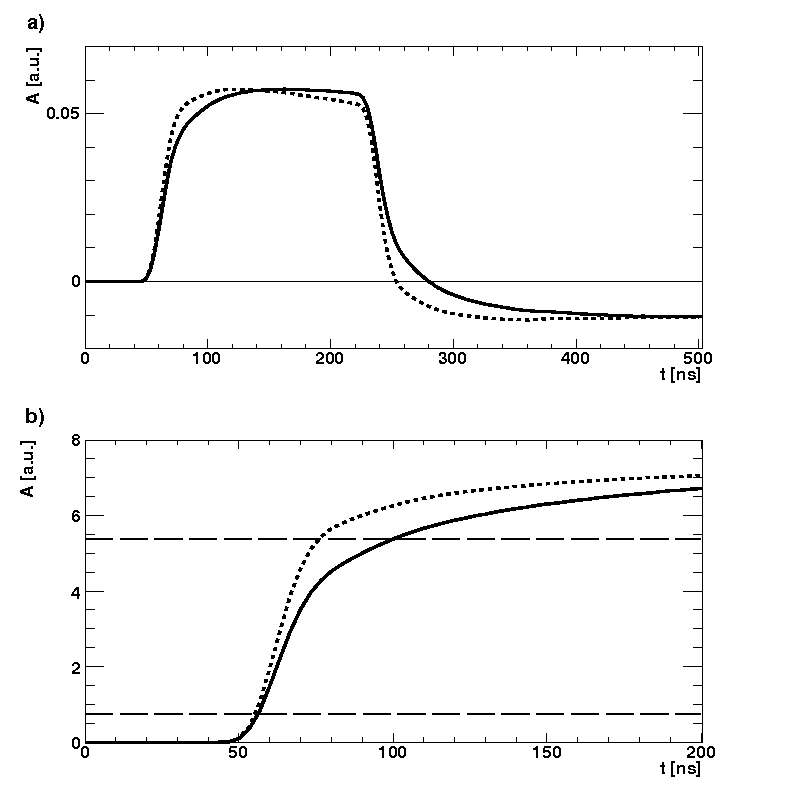}
  \caption{Processed average pulses from fig.~\ref{fig:ngamma}. a) Difference in ZCO time for a $\gamma$-ray pulse 
    (short-dashed) and a neutron (solid) pulse.
    b) Difference between the integrated rise
    time of a $\gamma$ ray (short-dashed) and a neutron (solid) pulse. 
    The points at 10 \% and 72 \% of the pulse height are 
    indicated by long-dashed lines. \label{fig:zerorise}
  }
\end{figure}

The filter $h(t)$ is itself composed of a convolution of three filters:
\begin{equation} \label{eq:ht}
 h(t) = h_{\mathrm{s}}(t) \ast h_{\mathrm{i}}(t) \ast h_{\mathrm{d}}(t).
\end{equation} 
The first filter, $h_{\mathrm{s}}(t)$, is a smoothing function, which
averages each sampling point with its neighbours within a time
interval $\tau_{\mathrm{s}}$ and the aim of which was to smooth out the
high-frequency noise of the signal. 
The integrating filter $h_{\mathrm{i}}(t)$ is a digital version of an analogue
integrator \cite[Ch. 16]{knoll}, which was obtained from the differential equation
\begin{equation} \label{eq:dp_out_dt}
  \frac{\sd p_{\mathrm{out}}}{\sd t} +
  \frac{p_{\mathrm{out}}}{\tau_{\mathrm{i}}} =  \frac{p_{\mathrm{in}}}{\tau_{\mathrm{i}}}
\end{equation} 
and which has the solution
\begin{equation} \label{eq:p_out}
  p_{\mathrm{out}} = \ex{-\frac{t}{\tau_{\mathrm{i}}}}
  \int \ex{ \frac{t}{\tau_{\mathrm{i}}}}
  \frac{p_{\mathrm{in}}}{\tau_{\mathrm{i}}}\sd t. 
\end{equation} 
Letting $p_{\mathrm{in}} = \delta(t)$, where $\delta(t)$ is the Dirac
delta function, and $\tau_{\mathrm{i}}p_{\mathrm{out}} = h_{\mathrm{i}}(t)$, the exponential integration filter
becomes
\begin{equation} \label{eq:hi}
 h_{\mathrm{i}}(t)=\ex{-\frac{t}{\tau_{\mathrm{i}}}}.
\end{equation}
Finally the pulse was differentiated with $h_{\mathrm{d}}(t)$, which was
a regular numerical differentiator filter, as described in \cite[Ch. 17]{knoll}. 
The filter $h(t)$ produces  
a bipolar pulse with a zero crossing, which depends on 
the type of particle that generated the pulse.
The selection of the time constants $\tau_{\mathrm{s}}$ and $\tau_{\mathrm{i}}$
was carefully examined and the optimal values for this setup was found
to be $\tau_{\mathrm{s}}=177$~ns and $\tau_{\mathrm{i}}=600$~ns. These values are similar to the time constants reported to be optimal for an analogue
discrimination system \cite{roush1964psd}. An example of a filtered neutron and $\gamma$-ray pulse is shown in fig.~\ref{fig:zerorise}a.

\subsection{ZCO: Integrated rise-time algorithm} \label{ss:irta}

By using digital pulse processing it is 
possible to evaluate the integrated rise time of the
pulse directly \cite{nima354_380}, instead of first shaping it to extract
the ZCO time, as explained in the previous subsection.
In the integrated rise-time algorithm (IRT) implemented in this work,
a mathematical integration filter
\begin{equation}
 h(t) = h_{\mathrm{i}}'(t) = \left\lbrace \begin{array}{ll}
                0, & t \leq 0 \\
                1, & t > 0\\
        \end{array}\right.
\end{equation} 
was used in eq.~(\ref{eq:ft}). The rise time of the integrated pulse was then extracted from the time 
difference between the position in time of the 10~\% and 72~\% values 
of the height of the integrated pulse. Its value depends on the type
of interacting particle, as illustrated in fig. \ref{fig:zerorise}b.

\subsection{CC: Slow component algorithm} \label{ss:sca}

\begin{figure}
 \begin{center}
\includegraphics[width=\columnwidth,bb=0 0 567 306]{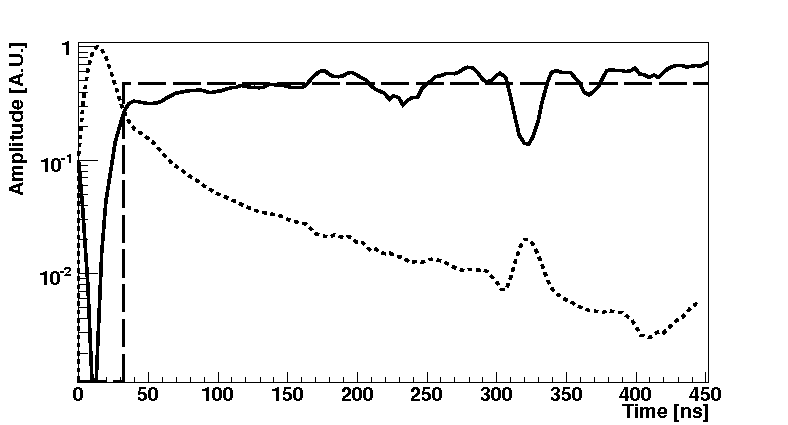}
\end{center}
\caption{Weighting function $w(t)$ of the GDM integral (solid) and slow component (dashed) algorithms shown together with an average neutron pulse (short-dashed).\label{fig:wt}}
\end{figure}

The slow component algorithm, based on the charge comparison method,
is a direct application of eq.~\ref{eq:psdobs}. The main assumption
of this algorithm is that all sampling points in the delayed part of
the pulse, corresponding to the slow component of the scintillation light, 
have equal weights. This was implemented by letting $w(t)$ be a step
function, with a constant value in the time range $t_1$ to $t_2$,
corresponding to the slow component, and 0 elsewhere. 
The parameters of the slow component algorithm are 
the start ($t=0$) and end ($t=T$) of the pulse and the times $t_1$ and
$t_2$, of which $t_1$ is most crucial, see fig.~\ref{fig:wt}.
The parameters
were determined in the following way.  
A digital LED, as described in subsection~\ref{ss:led}, was implemented and used for 
determination of $t=0$. The threshold of this
discriminator was set to a very low value of about $E_\mathrm{e}=5$~keV ($E_\mathrm{p}=150$~keV), that is lower
  than the hardware CFD threshold in
order not to loose any pulses.
For $t_2$ and $T$ an identical value of $533$~ns was chosen. For
times later than this value the pulse is barely above noise and contains
very little information. The chosen value of $t_1$ was $33$~ns, a
value which was obtained by careful optimization of the
figure-of-merit of the NGD (see subsection \ref{ss:fom}).

For the actual NGD based on the 
slow component algorithm, the value of the observable $S$ 
(see eq. \ref{eq:psdobs}) was evaluated as a function of the amplitude
of the input pulse (see section \ref{sec:analysis}).

The analogue version of the CC method is very similar
to the digital slow component algorithm described here. In the
analogue CC method the pulse is split and fed into two charge
sensitive ADC channels. In the ADC's the charges corresponding to the 
fast and slow components are integrated by applying suitable gates
at the times $t=0$ to $t=t_1$ and $t=t_1$ to $t=t_2$, respectively.
In practice the gates are usually set on the slow and total
components of the pulse.

\subsection{CC: GDM algorithm} \label{ss:gdm}

By taking advantage of the digital signal processing, one may choose
another, hopefully better, weighting function $w(t)$, than the simple
step function used in the slow component algorithm. The weighting
function can be evaluated directly from eq.~\ref{eq:optimalw}. This
was done by making use of the TOF parameter, by which neutrons and $\gamma$
rays also could be discriminated. Average $\gamma$-ray and neutron
pulses were created by selecting all pulses with a TOF value of
$\pm 1 \sigma$ from the maxima in the $\gamma$-ray and neutron TOF
distributions, respectively.
A total of about 70~000 $\gamma$-ray and 40~000 neutron pulses were
used for making the average pulses and these are shown in
fig.\ref{fig:ngamma}. 
The same digital leading edge discriminator and threshold
as in the slow component algorithm was used to define the start 
of the pulse ($t=0$). The length of the pulse was also the same,
i.e. $T=533$~ns. The created average pulses were then treated
in accordance with eq.~\ref{eq:optimalw} to get the optimal weighting
function $w(t)$, shown in fig.~\ref{fig:wt}. In the analysis, each sampling point of a pulse $p(t)$
was multiplied by the corresponding weight $w(t)$, and these products 
were summed together to form the observable $S$ according to
eq.~\ref{eq:optimalw}. The observable $S$ will be referred to as the 
GDM integral \cite{gattimartini}, $S_{\mathrm{GDM}}$. The GDM integral 
is usually normalized to the amplitude, $A$, of the pulse (see below).

\subsection{Figure-of-merit parameters} \label{ss:fom}

In order to quantify the results of the NGD, 
two different figure-of-merit (FOM) parameters were defined. One of
the most common definitions \cite{fom} of a FOM in this context is
\begin{equation} \label{eq:FOMdef}
  M=\frac{|X_\gamma-X_n|}{W_\gamma +
    W_n}.
\end{equation}
This is a unit-less ratio of the difference between the peak positions
$X_i$ divided by the sum of their FWHM, $W_i$ ($i=\gamma,n$). The
parameters $X_i$ and $W_i$ are obtained from a distribution of the
used NGD parameter, e.g. ZCO time, the
$S$ observable, or even TOF. An increased value of $M$ corresponds to a 
a better NGD.
The FOM parameter $M$ is usually determined by fitting two Gaussian or
similar functions to the distribution of the NGD parameter. 
\begin{figure} [ht] 
  \centering
\includegraphics[width=\columnwidth,bb=0 0 567 396]{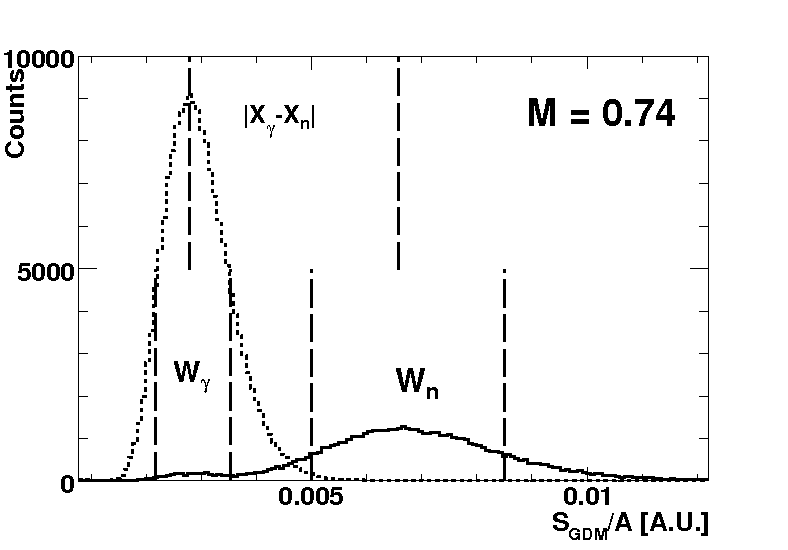}
  \caption{Illustration of the extraction of the
    parameters $X_{\gamma}, X_{n}, W_{\gamma}, W_{n}$, which are
    needed for evaluation of the FOM parameter $M$. 
    Neutrons (solid line) and $\gamma$ rays (short dashed line) were selected by making cuts on the TOF parameter (see text). The amplitude of the pulses shown in this figure are in the range 
    $E_\mathrm{e}=100$-$115$ keV for electrons, which corresponds to  
     $E_\mathrm{p}=760$-$830$ keV for recoil protons.
\label{fig:fomdef}
  } 
\end{figure}
The fitting method works well if the distributions have well defined
forms. This is usually not the case and therefore the following
method, which does not require any fitting, was developed and used.
Pulses due to neutrons and $\gamma$ rays were separated by making cuts
on the TOF parameter. The selected widths of the cuts were $\pm 2
\sigma$ around the maxima of the neutron and $\gamma$-ray TOF
distributions, which gave a clean enough selection.
Two separate distributions of the NGD 
parameter under evaluation were created, one with a TOF cut on neutrons,
the other with a TOF cut on $\gamma$ rays. These distributions had only a
single peak corresponding to neutrons and $\gamma$ rays,
respectively. The parameters $X_i$ and $W_i$ could then easily and
directly be extracted (without fitting) from the distributions as 
the maximum and FWHM of the distributions, respectively.
This procedure is illustrated in fig.~\ref{fig:fomdef} for NGD with the GDM algorithm.

The FOM parameter $M$ does not take into account any effects of bad
NGD due to random events or pile-up of
several close lying pulses in the liquid scintillator detector.
A mis-identification of neutrons as $\gamma$ rays, or vice versa, due
to such effects may be of great nuisance in real experiments, in
particular when the detector rates are high or when the ratio of 
number of detected neutrons to $\gamma$ rays is very small.
For example, an increase of the distance $|X_{\gamma}-X_i|$ between
peaks, which already are well separated in the distribution of an
NGD parameter, will increase $M$. 
However, the amount of neutrons mis-identified as $\gamma$ rays,
due to random and pile-up effects, will in this
case usually be the same even if $M$ has increased.

Another drawback of using the standard FOM parameter $M$ is that it
can only be extracted from one dimensional data, i.e. from a
distribution of just one NGD parameter.
In the analysis of data obtained in real experiments, correlations
between several NGD parameters are usually
evaluated. Another FOM parameter, which quantifies the
NGD, including random and pile-up effects,
and which can be used also in two dimensions, was therefore defined
as follows.

\begin{figure} [ht] 
  \centering
\includegraphics[width=\columnwidth,bb=0 0 567 680]{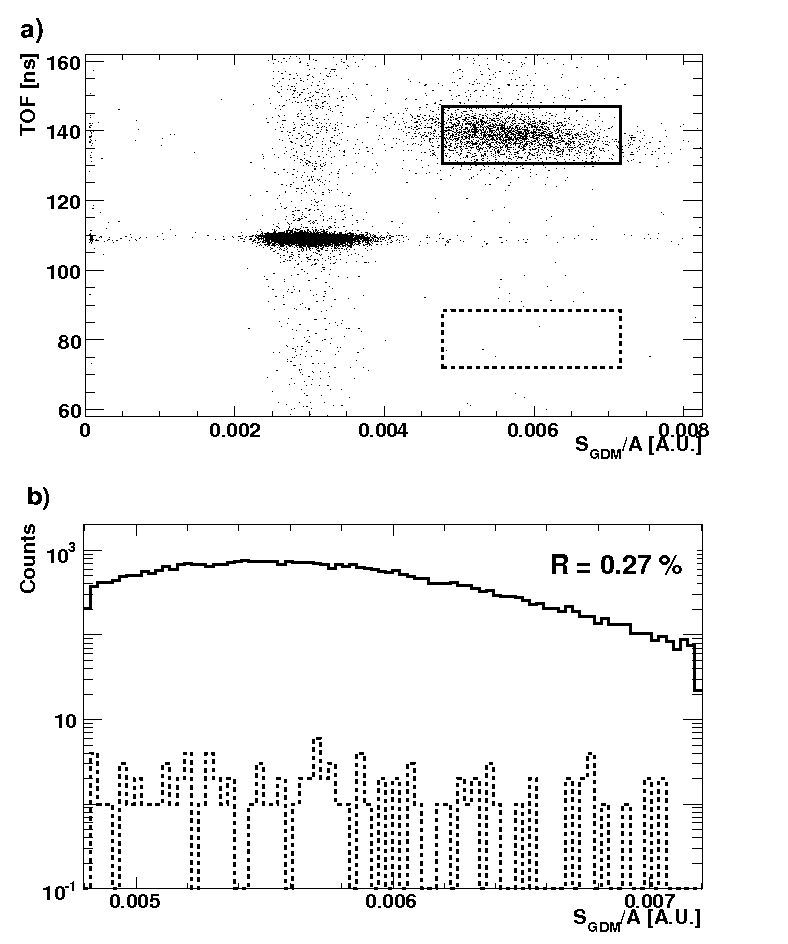}
  \caption{a) TOF versus $S_{\mathrm{GDM}}/A$ with gates on  
    neutron (rectangle with solid lines) and background 
    (dashed lines) events. b) Projection on the $S_{\mathrm{GDM}}/A$
    axis of the counts inside the solid and dashed rectangles in panel a).
    The amplitude of the pulses shown in this figure have an energy 
    in the range $E_{\mathrm{e}}=500$-$700$ keV for electrons, which corresponds to 
    $E_{\mathrm{p}}=2.2$-$2.7$ MeV for recoil protons.\label{fig:prob}
  } 
\end{figure}

Two dimensional (2D) cuts on the TOF parameter versus the evaluated NGD
parameter were applied. This is illustrated in fig.~\ref{fig:prob}a, which shows a 2D histogram of the TOF versus the $S_{\mathrm{GDM}}/A$ parameter for pulses of amplitude $E_{\mathrm{e}}=500$-$700$~keV.
In this plot, the centroid of the prompt $\gamma$-ray distribution,
which contains most of the events, is located at TOF $\sim
110$~ns and $S_{\mathrm{GDM}/A}\sim 0.003$. 
The neutron events are centred at TOF
$\sim 140$~ns and $S_{\mathrm{GDM}/A}\sim 0.0055$. Random TOF events, due to
the detection of $\gamma$ rays or neutrons from two different
$^{252}$Cf decays (or background $\gamma$ rays) in the BaF$_{2}$ and neutron detectors, are distributed anywhere parallel to the TOF
axis. Some of the random $\gamma$ rays can be seen as a vertical band
of counts centred at a $S_{\mathrm{GDM}}/A$ value of $\sim 0.003$.  Pile-up
events in the neutron detector may have any $S_{\mathrm{GDM}}/A$ value, which
is most clearly seen for the prompt $\gamma$ rays as the horizontal
band of counts centred at TOF $\sim 110$~ns. Simultaneous random and
pile-up events are homogeneously distributed anywhere in the 2D
histogram, including inside the 
neutron distribution. 

A 2D cut on the neutrons was set
as shown by the rectangle with solid lines in fig.\ref{fig:prob}a.
The width of the cut on the TOF parameter was chosen to
contain all counts within $\pm 2 \sigma$ from the maximum of the
neutron TOF distribution,
while the width of the cut on the evaluated NGD parameter
($S_{\mathrm{GDM}}/A$ in fig. \ref{fig:prob}) was one FWHM on the left side and two FWHM on the right side.
A 2D cut on background events, due to random coincidences and pile-up,
was selected as shown by the rectangle with dashed 
lines in fig.\ref{fig:prob}. For the NGD parameter the position of the 
background cut was identical to the position used for the neutron cut,
while for the TOF parameter it was located at the same distance from
the centroid of the prompt $\gamma$-ray distribution, but at a time before
instead of after this centroid. By this choice, it was assumed that 
the neutron cut on average contains as many background events as the
cut on background. Projections of the cuts on the $S_{\mathrm{GDM}}/A$ are
shown in fig.~\ref{fig:prob}b.
 
The new FOM parameter for quantifying the NGD was
defined as the ratio of the number of counts in the background ($N_\mathrm{b}$)
and neutron ($N_\mathrm{n}$) cuts:
\begin{equation} \label{eq:p_g}
R = \frac{N_\mathrm{b}}{N_\mathrm{n}-N_\mathrm{b}}.
\end{equation}
A small value of $R$ indicates a good discrimination of
real neutron events from any other events.
It should be stressed that the parameter $R$
only is useful for comparing how well different NGD algorithms 
can discriminate neutrons from $\gamma$ rays for analyses of the same
experimental data set. Two data sets with
differences in the relative number random or pile-up events will
give different values of $R$ even if the applied NGD
algorithm is identical.

\section{Analysis and results} \label{sec:analysis}

\subsection{Comparison of neutron-$\gamma$ discrimination
  algorithms} \label{ss:comp_ngd} 

\begin{figure*}[ht] 
 \begin{center}
\includegraphics[width=\textwidth,bb=0 0 567 240]{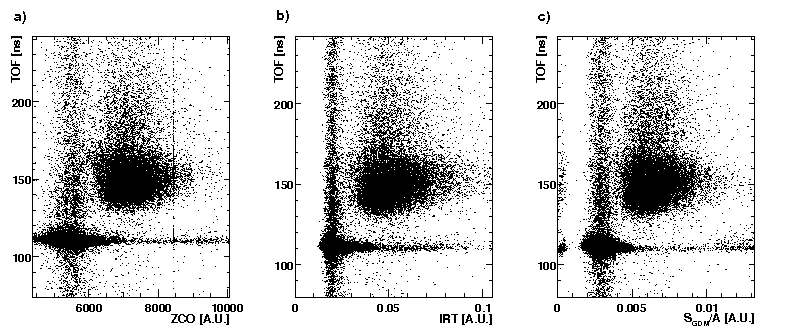}
\end{center}
\caption{Qualitative view of the neutron and $\gamma$-ray separation against TOF in the energy range $E_\mathrm{e}=90$-$700$ keV, $E_\mathrm{p}=730$-$2700$ keV for the:
         a) analogue separation,
         b) integrated rise time,
         c) GDM integral.\label{fig:tofs}}
\end{figure*}

\begin{figure*}[ht] 
 \begin{center}
\includegraphics[width=\textwidth,bb=0 0 567 208]{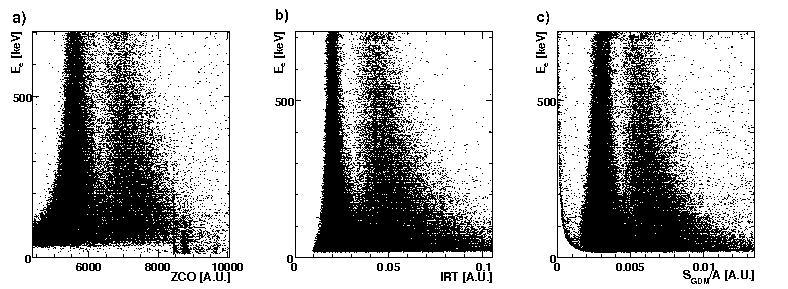}
\end{center}
\caption{Qualitative view of the neutron and $\gamma$-ray separation against $E_\mathrm{e}$ for the:
         a) analogue separation,
         b) integrated rise time,
         c) GDM integral.\label{fig:amps}}
\end{figure*}

A comparison of the NGD achieved with the digital PSD
algorithms and with the analogue setup are shown in figures
\ref{fig:tofs} and \ref{fig:amps}. Qualitatively the digital algorithms
show similar NGD properties as what was achieved with the analogue data.

\begin{figure*} [ht] 
  \centering
\includegraphics[width=\textwidth,bb=0 0 567 299]{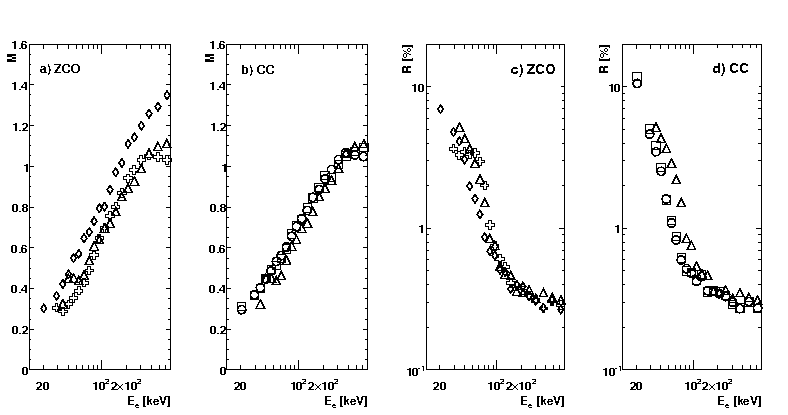}
  \caption{Figure-of-merit parameters for the four algorithms: convolution (crosses), integrated rise time (diamonds), slow component (circles) and GDM integral (squares), together with the analogue pulse shape discrimination (triangles).
      a) $M$ vs $E_\mathrm{e}$ for the two ZCO algorithms,
      b) $M$ vs $E_\mathrm{e}$ for the two CC algorithms,
      c) $R$ vs $E_\mathrm{e}$ for the two ZCO algorithms and
      d) $R$ vs $E_\mathrm{e}$ for the two CC algorithms.
      Statistical and systematical errors are about the same or smaller than the size of the symbols.
      \label{fig:fomglf}
  }
\end{figure*}

The FOM parameters $M$ and $R$ were evaluated for all digital 
algorithms and for the analogue data as a function of the amplitude of
the pulses. The obtained results are shown in fig.
\ref{fig:fomglf}. The expected increase of $M$ and decrease of $R$ 
with increasing amplitude is clearly visible in the figure.
In general, the differences in the obtained FOM parameters 
for the different algorithms, are rather small. All digital algorithms give at least as good or better FOM parameter than was obtained with the analogue data. The IRT 
algorithm based on the ZCO method gives the best $M$ values (see fig. \ref{fig:fomglf}a).
The convolution algorithm gives somewhat worse values of both $M$ and $R$
(fig. \ref{fig:fomglf}a and c) for pulses with amplitudes in the energy
range $E_\mathrm{e} \simeq 40$-$100$ keV.

\subsection{ADC sampling frequency and bit resolution} \label{ss:adc}

A detailed investigation of the performance of the IRT
(based on the ZCO method) and GDM (CC method) algorithms 
as a function of the sampling frequency and bit resolution
of the ADC was performed as described in this subsection. 

From the gain and time matched event-by-event data new data sets were
created with a reduction of either the bit resolution or the sampling
frequency. The new data sets were then analysed with the same code as
what was used for the original data set.

\begin{figure*} [ht] 
  \centering
\includegraphics[width=\textwidth,bb=0 0 567 268]{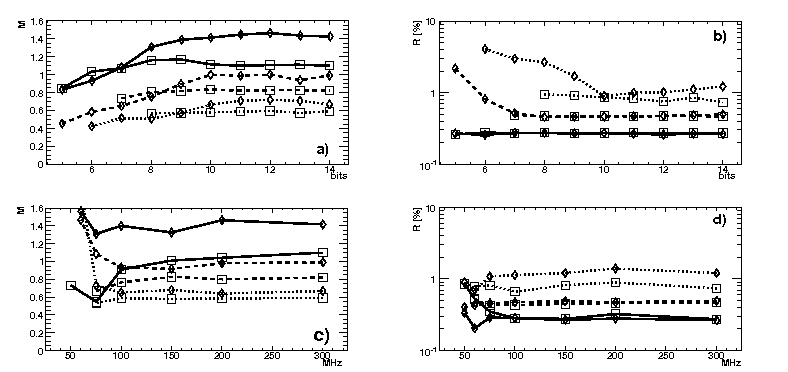}
  \caption{The FOM parameters $M$ (panel a and c) and $R$ (b, d)
    as a function of ADC bit resolution (a, b) and sampling
    frequency (c, d). Results are
    shown for the algorithms of type IRT (diamonds; ZCO  method) and GDM (squares; CC
    method). Pulses in three different energy ranges are shown: 
    $E_\mathrm{e}=50$-$70$ keV, $E_\mathrm{p}=500$-$540$ keV (dotted),
    $E_\mathrm{e}=115$-$135$ keV, $E_\mathrm{p}=830$-$920$ keV (dashed),
    $E_\mathrm{e}=500$- $700$ keV, $E_\mathrm{p}=2200$-$2700$ keV (solid).
    Statistical and systematical errors are about the same or smaller than the size of the symbols.\label{fig:fombitfreq}
  }
\end{figure*}

The bit reduction was achieved by performing an integer division by
$2^{14-b}$ of the value of each sampling point value, where $b$ is the
number of bits in the new data set. The resulting value was then
multiplied by the same factor. This made it possible to use exactly
the same code for the analysis of all different data sets, because 
there was no need for corrections like rescaling of threshold values,
etc. A total of 10 data sets with bit resolutions from 5 to 14 bits were
created.
The sampling frequency was lowered by removing a number of data points
from the 300 Ms/s data set, resulting in new data sets corresponding
to sampling frequencies of 150, 100, 75, 60 and 50 Ms/s. In addition a
data set with 200 Ms/s was available from the measurement 
with the analogue PSD unit as  described in section
\ref{sec:experiment}. Thus, a total of 7 data sets with samplings
frequencies from 50 to 300 Ms/s, were created.
The results obtained using the IRT and GDM algorithms for pulses of
three different amplitude ranges 
are presented in fig. \ref{fig:fombitfreq}.

As seen in fig. \ref{fig:fombitfreq}a, both algorithms
show a clear saturation of the FOM parameter $M$ when the bit
resolution is increased above a certain value. The
point of saturation is almost independent of the amplitude of the
pulses and occurs at about 9 bits for the GDM algorithm and
at about 10 bits for the IRT algorithm.
The minimum required bit resolution depends, however, strongly on the
requested dynamic range of the amplitudes of the pulses. In this work
the dynamic range was $E_\mathrm{e}=15$-$700$ keV ($E_\mathrm{p}=250$-$2700$ keV),
which is much smaller than what commonly is used in real experiments.
A rescaling of the dynamic range from the one presently used
to more reasonable values is shown in table \ref{tab:energylimits}.
A dynamic range corresponding to a maximum energy deposit of
$E_\mathrm{e}=5.6$~MeV  ($E_\mathrm{p}=12$~MeV), which requires a bit resolution of 12 bits,
seems adequate in most experiments.

\begin{table} [ht] 
  \caption{The dependency of the maximum electron energy $E_\mathrm{e}$ and
    corresponding recoil proton energy $E_\mathrm{p}$ on 
    the ADC bit resolution, with retained saturated (best) value of the FOM
    parameter $M$ shown in fig.~\ref{fig:fombitfreq}a\label{tab:energylimits}.
  }
  \begin{tabular*}{\columnwidth}{@{\extracolsep{\fill}}ccccc}
    \hline
    Nr of bits & 
    \multicolumn{2}{c}{Integrated rise time  (ZCO)} &
    \multicolumn{2}{c}{GDM integral (CC)} \\
    & $E_\mathrm{e}$ & $E_\mathrm{p}$ &  $E_\mathrm{e}$ & $E_\mathrm{p}$ \\
    & [MeV] & [MeV] & [MeV] & [MeV] \\
    \hline
    9  &  0.70  &   2.7    &  0.35      & 1.7\\
    10 &  1.4  &    4.4   &    0.70    & 2.7\\
    11 &  2.8  &    7.3   &   1.4     & 4.4\\
    12 &  5.6  &    12   &    2.8    & 7.3\\
    13 &   11 &    -   &    5.6    & 12\\
    14 &   22 &    -   &    11    & -\\
    \hline
  \end{tabular*}
\end{table}

The dependency of the FOM parameter $R$ on the bit resolution is shown 
in fig. \ref{fig:fombitfreq}b. The two algorithms give
very similar results for the intermediate and high energy pulses,
while the IRT algorithm gives slightly better 
$R$ values for the low energy pulses.
In contrast to the results obtained for the $M$ parameter (panel a),
there is an energy dependency of the bit resolution value at which the 
 $R$ parameter saturates. For the low energy pulses, $R$ saturates at
a bit resolution of $\ge 9$ bits and $\ge 10$ bits for the IRT 
and GDM algorithms, respectively. For the intermediate
energy pulses both algorithms saturate at $\ge 7$ bits, while no
saturation point was observed for the high energy pulses for bit
resolutions of 5 bits and higher. No $M$ and $R$ values could be obtained for the small and intermediate amplitude pulses at the lowest bit resolutions.

The results obtained for different sampling frequencies are shown in
fig. \ref{fig:fombitfreq}c and d, regarding the $M$ and $R$
parameters, respectively. In general one can say that there is no
strong dependency of $M$ and $R$ on sampling frequency. The $M$
values saturate and become more or less constant above 100 Ms/s,
except for the high energy pulses treated by the GDM 
algorithm, in which case $M$ increases slowly above 100 Ms/s. No $M$ values could be evaluated 
for the intermediate and low energy pulses with the
IRT algorithm below 75 Ms/s, which apparently does
not work so well below this sampling frequency for pulses with 
smaller signal to noise ratio.
The $R$ values saturate for sampling frequencies already above about
75 Ms/s. The apparently odd frequency behaviour for low-energy pulses is due to asymmetries in the neutron and $\gamma$-rays distributions and does not imply a better NGD.

\subsection{Time resolution at 100 Ms/s}

\begin{figure} [ht] 
  \centering
\begin{center}
\includegraphics[width=\columnwidth,bb=0 0 567 243]{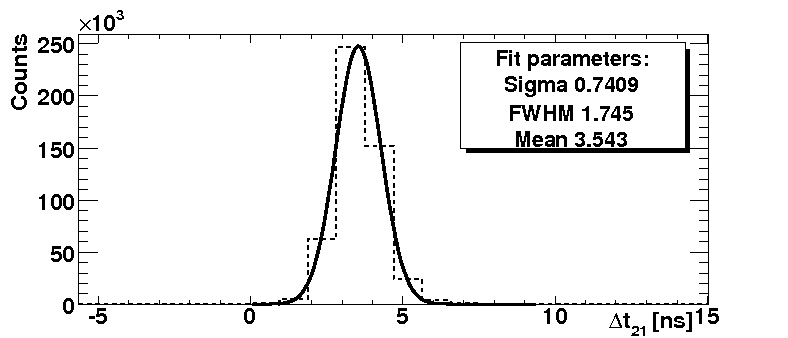}
\end{center}
  \caption{Distribution of the time difference extracted from two waveforms recorded at 100 Ms/s. See text for details.\label{fig:t21}
  } 
\end{figure}

In a fully digital NGD system an external TAC, as used in this work, will not be available. For the TOF measurement, the starting time of the pulse in the neutron detector must instead be determined from the digitized waveform itself. A test of the influence of the finite sampling frequency on the achievable time resolution was made in the following way. A timing parameter was extracted from the digitized waveforms of the TNT2 channels 1 and 2, each recorded with a sampling frequency of 100 Ms/s. The waveforms were integrated as described in subsection~\ref{ss:irta} and a LED (see subsection~\ref{ss:led}) with a threshold set at half of the amplitude of the integrated pulse, was applied. Pulses of all amplitudes were used in the analysis. The timing parameter $\Delta t_{21}$ was created by taking the difference between the extracted LED times of each of the two channels. The distribution of $\Delta t_{21}$ is shown in fig.~\ref{fig:t21}. The FWHM of the distribution is 1.7 ns, which is the contribution of the finite sampling frequency (100 Ms/s in this case), to the total FWHM. The achievable intrinsic time resolution of a liquid scintillator detector plus PMT is typically FWHM~=~1.5~ns or larger. With the present data it was not possible to perform the analysis at higher sampling frequencies, but it is expected that the FWHM will decrease linearly with increasing sampling frequency. Thus, already at 200~Ms/s the contribution of the finite sampling frequency to the total FWHM of the time resolution is almost negligible.

\section{Summary and conclusions} \label{sec:con_sum}

In this work four different digital pulse-processing algorithms for
discrimination of neutrons and $\gamma$ rays in a liquid scintillator
detector have been developed and compared to each other and to data
obtained with an analogue neutron-$\gamma$ discrimination unit. Two of
the digital algorithms were based on the charge comparison method,
while the analogue unit and the other two digital algorithms were
based on the zero-crossover method. Two different figure-of-merit
parameters, which quantifies the neutron-$\gamma$ discrimination
properties, were evaluated. All of the digital algorithms gave similar or
better figure-of-merit values than what was obtained with the
analogue setup.

A detailed study of the discrimination properties as a function of
sampling frequency and bit resolution of the ADC was
performed. The general conclusion is that
an ADC with a bit resolution of 12 bits and a sampling
frequency of 100 Ms/s is adequate for achieving an optimal
neutron-$\gamma$ discrimination for pulses having a dynamic energy
range of 0.02 - 5.6 MeV and 0.3 - 12 MeV for $\gamma$ rays and neutrons, respectively. The influence of a finite sampling frequency on the time resolution was also investigated. A FWHM of 1.7~ns was obtained at 100~Ms/s.

In order to further increase the quality of the neutron-$\gamma$ discrimination
it is necessary to handle random and pileup effects 
in an adequate way. Such effects will become more problematic
in experiments using very high-intensity stable or high-intensity
radioactive beams, in which an increased $\gamma$-ray background
radiation is present. For such experiments it will be necessary to
further develop the digital pulse-shape algorithms for rejection or recovery of pulses which are distorted

\section*{Acknowledgements}

We are grateful to J. Ljungvall for the TNT2 library and to H. Mach for supplying the BaF$_2$ detector. This work was partially supported by the Swedish Research Council.

\bibliographystyle{elsart-num}

\begin{thebibliography}{10}
\bibitem{TNT2}
{The TNT2 was designed and built by P. Medina, \etal\ at IPHC, Strasbourg.
  Presently it is available commercially from CAEN as model N1728.}
\end{thebibliography}

\end{document}